\documentclass[manuscript,screen,nonacm]{acmart}
\makeatletter
\def\@ACM@checkaffil{
    \if@ACM@countrypresent\else
        \ClassWarningNoLine{\@classname}{No country present for an affiliation}%
    \fi
}
\makeatother
\acmConference[]{} 

\newcommand{\workshopname}{GenAICHI: CHI 2024 Workshop on Generative AI and HCI}
\newcommand{\licensedetails}{Licensed under a Creative Commons Attribution 4.0 International License (CC BY 4.0). Copyright remains with the author(s).}
\newcommand\extrafootertext[1]{
    \bgroup
    \renewcommand\thefootnote{\fnsymbol{footnote}}%
    \renewcommand\thempfootnote{\fnsymbol{mpfootnote}}%
    \footnotetext[0]{#1}%
    \egroup
}

\AtBeginDocument{ 
    \fancypagestyle{firstpagestyle}{
        \fancyhf{}
        \fancyfoot[L]{\sffamily\footnotesize \workshopname}%
        \fancyfoot[C]{\sffamily\footnotesize \thepage}
    }
    \fancyhf{}
    \fancyhead[L]{\sffamily\footnotesize\shorttitle}
    \fancyhead[R]{\sffamily\footnotesize\shortauthors}
    \fancyfoot[L]{\sffamily\footnotesize\workshopname}%
    \fancyfoot[C]{\sffamily\footnotesize\thepage}
    \extrafootertext{\licensedetails}
}

\usepackage{colortbl,subcaption}

\begin{document}

\title[Co-creativity for AI Storytelling]{Holding the Line: A Study of Writers’ Attitudes on Co-creativity with AI}

\author{Morteza Behrooz}
\authornote{Email: \texttt{morti@meta.com}}
\author{Yuandong Tian}
\author{William Ngan}
\author{Yael Yungster}
\affiliation{
  \institution{FAIR, Meta}
}
\author{Justin Wong}
\affiliation{
  \institution{University of California, Berkeley}
}
\author{David Zax}

\begin{abstract}
 Generative AI has put many professional writers on the defensive; a major negotiation point of the recent Writers Guild of America’s strike concerned use of AI. However, must AI threaten writers, their livelihoods or their creativity? And under what conditions, if any, might AI assistance be invited by different types of writers (from the amateur to the professional, from the screenwriter to the novelist)? To explore these questions, we conducted a qualitative study with 37 writers. We found that most writing occurs across five stages and within one of three modes; we additionally map openness to AI assistance to each intersecting stage-mode. We found that most writers were interested in AI assistance to some degree, but some writers felt drawing firm boundaries with an AI was key to their comfort using such systems. Designers can leverage these insights to build agency-respecting AI products for writers.
\end{abstract}


\maketitle

 \section{Introduction and Background}

 As AI becomes a topic of interest for various societal groups, AI interaction models play a crucial role in shaping the public perception of AI. It could be argued, for instance, that ``prompt engineering'' fosters a view of AI that is more about end-to-end automation, rather than a processes of creativity or collaboration, where humans reflect, refine, and form thoughts. While full automation is sometimes desirable, it may not always be the optimal paradigm, and exploring other interaction models help cement the perception of AI as a potentially collaborative, co-creative force that can be congruent with the human cognitive \textit{processes}. 
 
 In this paper, we focus on exploring co-creativity in AI story generation. By ``story'', we refer to a broad set of writing and creative tasks that focus on narrative construction, activities ranging from screenwriting to fiction writing to the designing of Dungeons \& Dragons campaigns. By ``story generation'' we mean the computational generation of various types of stories \cite{storygenSurvey1, storygenSurvey2, storygenSurvey3} which is also a focus of LLM use recently; and by ``co-creativity'' \cite{rezwana2022designing, ccmain2, kantosalo2020modalities}, we broadly refer to a process of a human collaborating with an AI system in order to facilitate and improve the process of story creation. 
 
 While previous work covers the theoretical aspects of co-creativity interactions \cite{kantosalo2020modalities,beyondPrompts}, it can be distant from the communities it intends to address in specific domains, as argued in \cite{otherCCqualstudy}. Recent work also focuses on creating interaction models motivated by LLM capabilities \cite{TaleBrush, g-wordcraft}, without due focus on interaction itself. Our work views the co-creative AI interaction from the perspective of writers and find principles that can inspire both design and AI research, complementing recent other work \cite{unmet} on finding unmet needs of writers by adding qualitative reports and analysis. We also hope to showcase, through an example, how the ``necessary automation'' perception of AI is consequential, influenced by HCI, and at times, unnecessary.
  
\section{A Qualitative Study on Writers}

  Our broad research questions, informed by the above goal, were:
  
  \begin{itemize}
  \setlength\itemsep{0em}
      \item What are the constituent parts and processes in creative writing? What are the user needs in each? 
      \item Which parts and processes are more appealing for AI co-creativity and why? 
      \item What design principles and LLM capabilities are needed for AI co-creativity in story writing?
  \end{itemize}

 \subsection{Participants, Protocol and Procedure}

  We recruited a diverse mix of participants who engaged in a wide range of writing types (novels, memoir, screenplays, game-related writing, diaries and blogs, fiction, non-fiction, etc.), as well as a mix of years of experience writing. The final set of participants included 37 individuals (18 of whom identified as female and 19 of whom identified as male). The median age was 42, with a min of 23 and max of 70. All study sessions were remote video calls on a computer screen. Study sessions were each 90 minutes long; compensation was a \$75 gift card. Each session consisted of 3 segments:
 
 \begin{itemize}
   \setlength \itemsep{0em} 
      \item A \textbf{semi-structured qualitative interview} segment. We explored the writers’ writing process and asked all participants to try ChatGPT capabilities in storytelling by directing them to perform a relevant task of their choice (e.g. generate a short story outline, get feedback on an outline, etc.).
     \item We followed the interview with a \textbf{concept testing phase}, in which we showed a series of static design sketches (See the Appendix) as stimuli, where the AI's help in the storytelling process would take a different shape than generating story segments through prompting (or prompts delivered through dialogue).
     \item Lastly, in a conversational \textbf{participatory design} exercise, we asked participants about specific features they would see as useful and the place of AI in their personal writing practice.

 \end{itemize}

 \section{Study Findings}
 
  Through our interviews and analysis, we identified five writing stages (often corresponding to the phase of a project) and three writing modes (often corresponding to the writer’s overall attitude or purpose related to a given project). Stages and modes appear differentially correlated to a writer’s openness to AI assistance. 
 
 \subsection{{Writing stages have to be explicitly recognized}} We found at least 5 major stages that most participants went through in the process of writing stories. While neither exhaustive nor perfectly sequential, these stages point to a changing focus and task dynamic during writing. 

 \begin{itemize}
     \item \textit{Ideation}: Entertaining the core ideas in a story they may write;
     \item \textit{Drafting}: Starting somewhere in the story to write in text, e.g. a beginning, the plot, or a scene;
     \item \textit{Story management}: Taking various forms across writing genres, this stage involves indexing ideas, keeping track of characters, features, details, artifacts, doing research, etc.
     \item \textit{Feedback}: Getting feedback from a trusted partner or reader about on an early draft; and,
     \item \textit{Revision}: Iterative refinement of all aspects of the writing.
 \end{itemize}

  \subsection{{Writing modes have to be recognized to effectively engage in co-creativity}} 
  
  We also found three main emergent modes of writing, which we named \textit{hedonic}, \textit{scribe}, and \textit{artiste}. In \textit{hedonic} mode, writers are more free and casual in their writing, and prioritize entertainment and potentially learning. We found younger, budding writers, or those who said they dabbled in writing as a hobby, to be more likely to report being in this ``hedonic'' mode. In \textit{scribe} mode, writers have a more specific goal, and more structured evaluation and benchmarks, and potentially a deadline; e.g. within a commercial context. In this mode, often the writing product is being generated primarily for an end (like earning wages) other than pleasure. In \textit{artiste} mode, the focus is on the joy of authentic self-expression, deep emotional exploration, and perfection or innovation within a learned craft of writing. 
  
  \subsection{{Modes and stages can help predict a writer's degree of openness to AI co-creativity}} 
  
  \label{sec:sm} Participants expressed varying degrees of openness to engaging in co-creativity with an AI system or agent, depending on which stage or mode of writing they are in. Table~\ref{tab:AIopenness} shows a relative openness matrix.

  \begin{table}[!h]
    \centering
    \caption{Openness to AI in different stage-modes of writing. \checkmark denotes significant openness, X denotes significant reservation, and \textit{partially} denotes a combination where most participants were somewhat open to AI's help, but either conditionally or partially.}
    \label{tab:AIopenness}
    \begin{tabular}{|c|c|c|c|c|c|}
    \hline
    \textbf{} & \textbf{Ideation} & \textbf{Drafting} & \textbf{Story Management} & \textbf{Feedback} & \textbf{Revision} \\
    \hline
    \textit{\textbf{Hedonic}} & \cellcolor{green!25}\checkmark & \cellcolor{green!25} \checkmark & \cellcolor{green!25}\checkmark & \cellcolor{green!25}\checkmark & \cellcolor{green!25}\checkmark \\
    \hline
    \textit{\textbf{Scribe}} & \cellcolor{green!25}\checkmark & \cellcolor{yellow!25}\textit{partially} & \cellcolor{green!25}\checkmark & \cellcolor{green!25}\checkmark & \cellcolor{yellow!25}\textit{partially} \\
    \hline
    \textit{\textbf{Artiste}} & \cellcolor{yellow!25}\textit{partially} & \cellcolor{red!25}X & \cellcolor{green!25}\checkmark & \cellcolor{green!25}\checkmark & \cellcolor{red!25}X \\
    \hline
    \end{tabular}
  \end{table}
    
  To deeply explore every intersecting stage-mode (all 15 cells in the chart) is beyond the scope of this particular paper; however, in the following, we will observe a few  patterns in the results and draw a few conclusions. First, a writer in a ``hedonic'' mode (the top row in Table \ref{tab:AIopenness}) is often more open to AI assistance across all stages of the work. Our results suggest two reasons: 1) the writer in the ``hedonic'' mode is often a student who finds AI assistance as a means of learning, and, 2) the writer is often not yet sufficiently identified with the craft of writing to feel like a co-creative AI is usurping their own creativity. The writer in the ``artiste'' mode, by contrast (the bottom row in Table \ref{tab:AIopenness}), is the most resistant to AI assistance, or at least, the most \textit{particular} and \textit{selective}. In this mode, the writer is likely to have invested countless hours in honing their craft, and having developed specific personal cognitive processes to achieve creative goals, leading them to be more identified with their writing and more reluctant to outsourcing components of it. Similar to an athlete wanting to win a contest honestly, the writer in the artiste mode often feels that her own achievement might be diminished by the assistance of AI. Said one artiste-leaning writer, ``\textit{Why would I want to use this? This is like asking a trained NASCAR driver if they’d want to use a self-driving car?}'' Thus, artistes also gain high emotional rewards fro engaging in their activity as a hobby. 

  Interestingly, though, most writing modes – even the ``artiste'' mode – offer high openness to AI assistance during the two particular writing phases of 1) Story Management and 2) Feedback. In the case of Story Management, many writers saw AI's assistance as simply an extension of the writing software they often already used to manage complex projects (e.g. the outlining and organization software popular with novelists, Scrivener). In the case of Feedback, writers saw value in AI's supposed \textit{objectivity} in assessing a work; many writers reported challenges in assessing the complex emotional motivations behind the feedback they received from human beings.

\subsection{Design Challenges Presented by the Stage-Mode Framework}

  Designing for the hedonic mode may be relatively easy. Those in a hedonic mode – particularly students and early-career explorers, or those without much study of their craft undergirding their writing practice – tended to see almost all interventions and suggestions from an AI as value add. It is in the central and lower rows of our stage-mode framework (e.g. artiste mode) that a significant design challenge is presented. The writer in the artiste mode has very low error tolerances for suggestions from AI and has a very high bar for what is considered an appropriate and helpful intervention or suggestion from AI. Especially relevant to this mode, many writers imagined AI help to be completely different than AI drafting any words that breaks their creative flow, and focused on the \textit{process}, e.g. by letting the writer ``talk to their own story's character'' to get out of creative block.

  At the heart of this is a desire of the writer in artiste mode to protect her own agency and creativity –– to have it occasionally supported, but never usurped. Writers who tended to operate in the artiste mode –– often experienced novelists and screenwriters –– expressed a concern around \textbf{losing agency, ownership, and expression} \cite{ownershipCC1, ownershipCC2}. A related area of concern was about the \textbf{external perception of ownership}, and whether their work would be seen by the world as fully theirs. At least one writer we spoke to wanted to wait until norms around AI use had been widely discussed and adopted in the professional writing community, before they decided ``how much AI'' to use. 

  Writers were also concerned about stories losing the ``human touch''. Many writers who said they were open to collaborating with another human being (e.g. a friend) were not as readily open to collaborating with a machine, even if the process might be smoother. One reason had to do with their understanding of how LLM’s had been created and how they functioned; it felt meaningful to them that when collaborating with another human being, ideas generated by the human were often based on real, tangible, lived experience – rather than vast patterns from an enormous dataset. To bring something so mechanized into their writing process felt anathema; however, it is probable that the first users of typewriters and word processors felt something similar. Here, offering AI explainability, in-situ education of how LLMs work, and exploring characterful agents may be interesting.

\subsection{The Need for Effective Communication of Boundaries with the AI}

  In exploring AI with writers who tended to the artiste mode, an interesting pattern emerged. Many of these writers were reflexively dismissive of AI, only because they initially \textit{struggled to imagine} a version of AI that might be sensitive enough to help preserve and affirm their agency as humans (pointing to their perception). Asked to imagine an AI that, in the manner of a sensitive and emotionally intelligent writer’s assistant, was extremely cautious about when and how to offer assistance, many writers who skewed artiste expressed openness to the idea of being supported by AI.

  A recurrent theme in participants' reaction to the ideas and stimuli presented was their strong desire for the ability to direct and set interaction \textit{boundaries} with a co-creative AI in story writing. In some cases, a writer was open to working alongside an AI if the AI ``stayed in their lane'' and did not volunteer creative ideas unless ``directly prompted to do so''. Notably, this desire for control over boundaries was even strong within less-experienced writers or those who mostly used hedonic writing modes. Even P3, a young writer who mostly operated in the hedonic mode, said that they would want to ``block, limit, and define how the AI would interact'' with them. Several writers further said they wanted the ability to specify \textit{how} the AI should generate text or ideas: For instance, some writers were open to \textit{concrete and specific} ideas from an AI, while other writers were only open to more \textit{abstract and general} ideas that the user could then further articulate, steer, and specify. This point to a need for personalized and negotiated collaborative interactions.

\section{Conclusions, Limitations and Future Work}

In this paper, we describe our findings of a study on a diverse set of story writers and their attitude towards AI co-creativity. We find stages and modes of writing where intersecting stage-modes show differential openness to AI assistance and need particular design solutions. Our study was geographically limited to the U.S. and did not include an interactive experience. While we continue to analyse our results for more specific recommendations, addressing those limitations, along with conducting studies in other domains of creativity may yield more design recommendations.

\bibliographystyle{ACM-Reference-Format}
\bibliography{bib}

\newpage
\section{Appendix}
\label{appendix}

 \begin{figure}[!h]
     \centering
     \begin{subfigure}{0.3\textwidth}
         \centering
         \includegraphics[width=\textwidth]{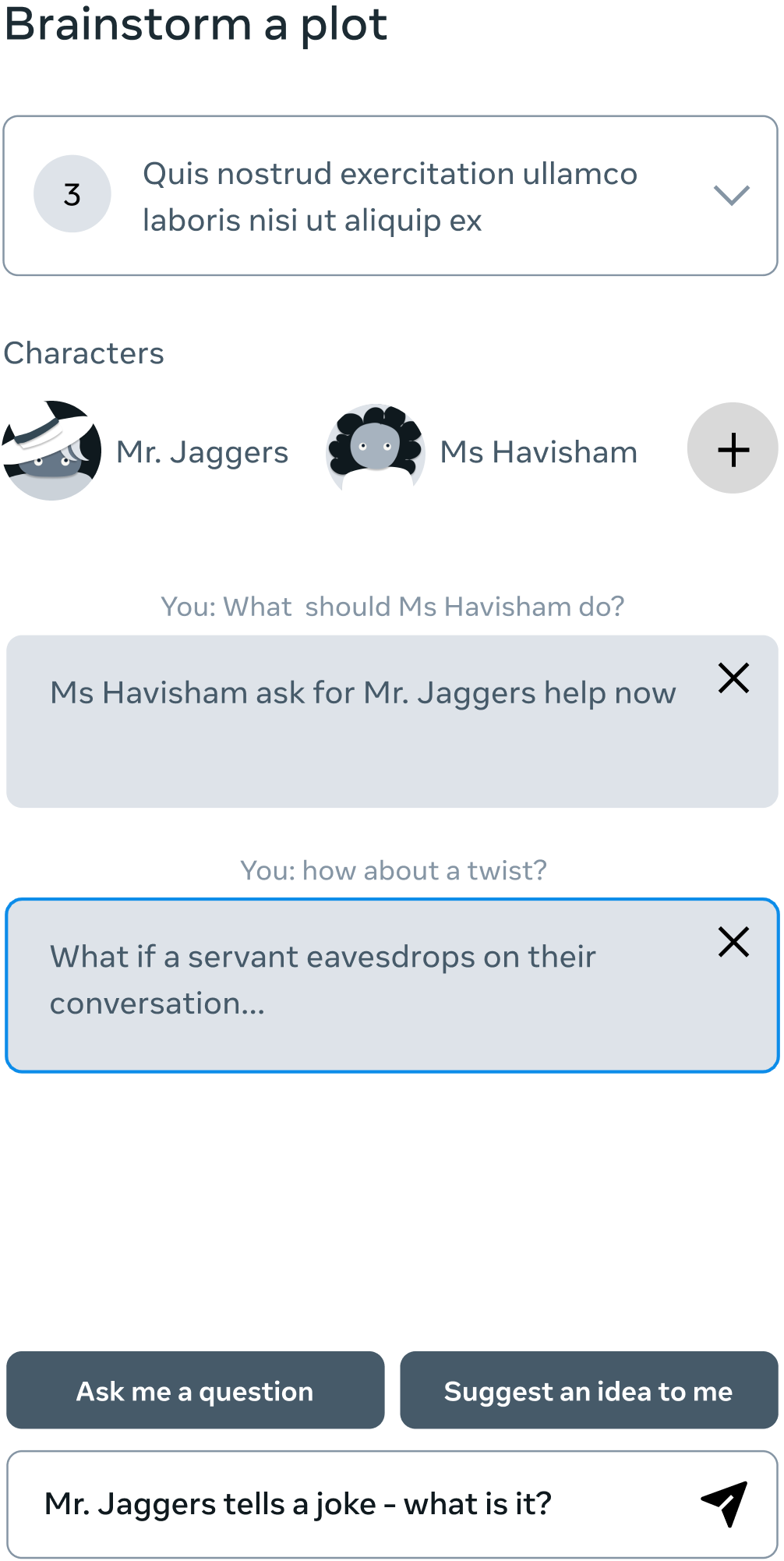}
         \caption{A design idea illustrating a brainstorming with AI about a plot.}
         \label{fig:1.0}
     \end{subfigure}
     \hfill
     \begin{subfigure}{0.3\textwidth}
         \centering
         \includegraphics[width=\textwidth]{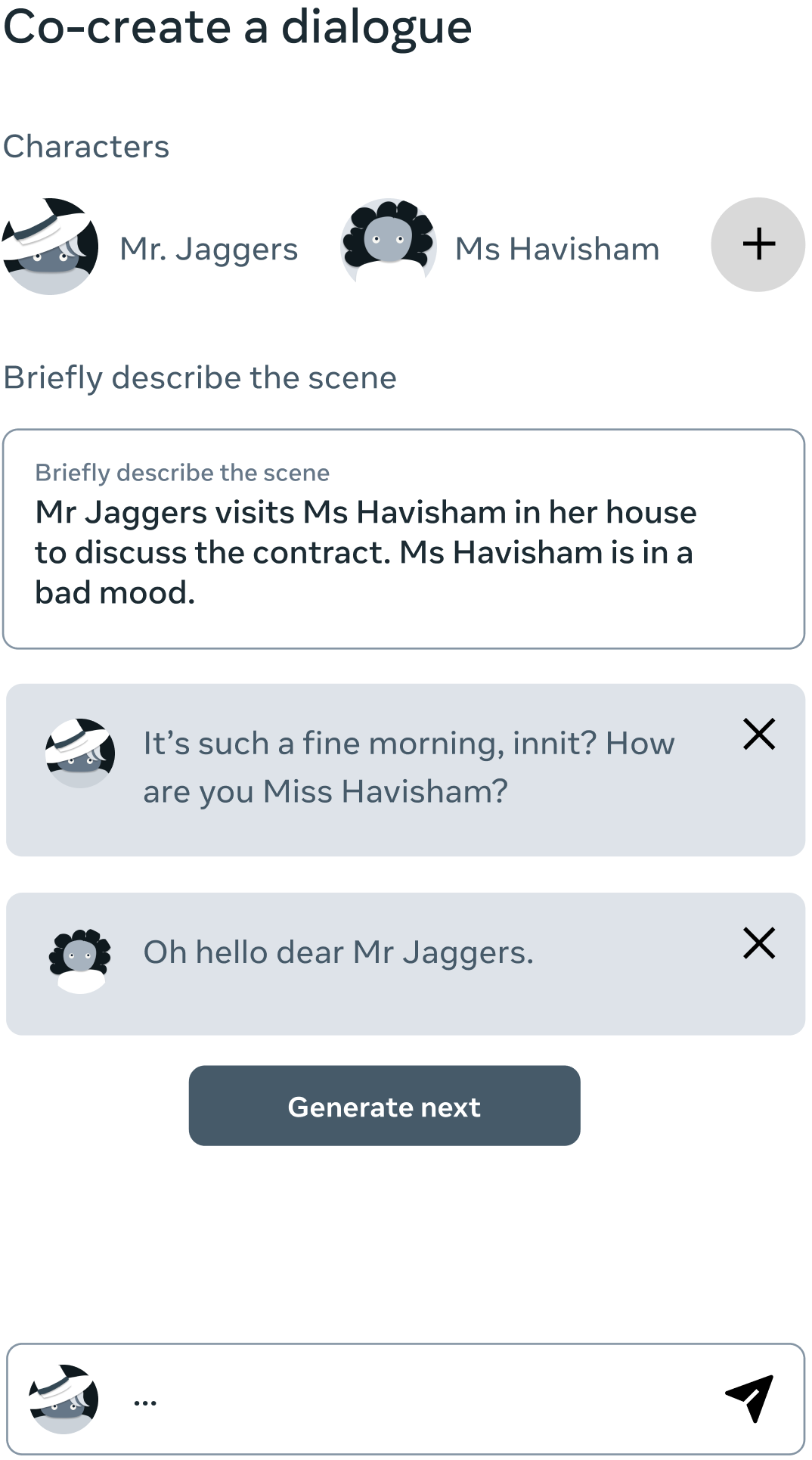}
         \caption{A design idea illustrating the AI helping the writer immerse herself in the story by allowing characters to have a dialogue, or optionally, use the dialogue in a resulting story.}
         \label{fig:1.1}
     \end{subfigure}
     \hfill
     \begin{subfigure}{0.3\textwidth}
         \centering
         \includegraphics[width=\textwidth]{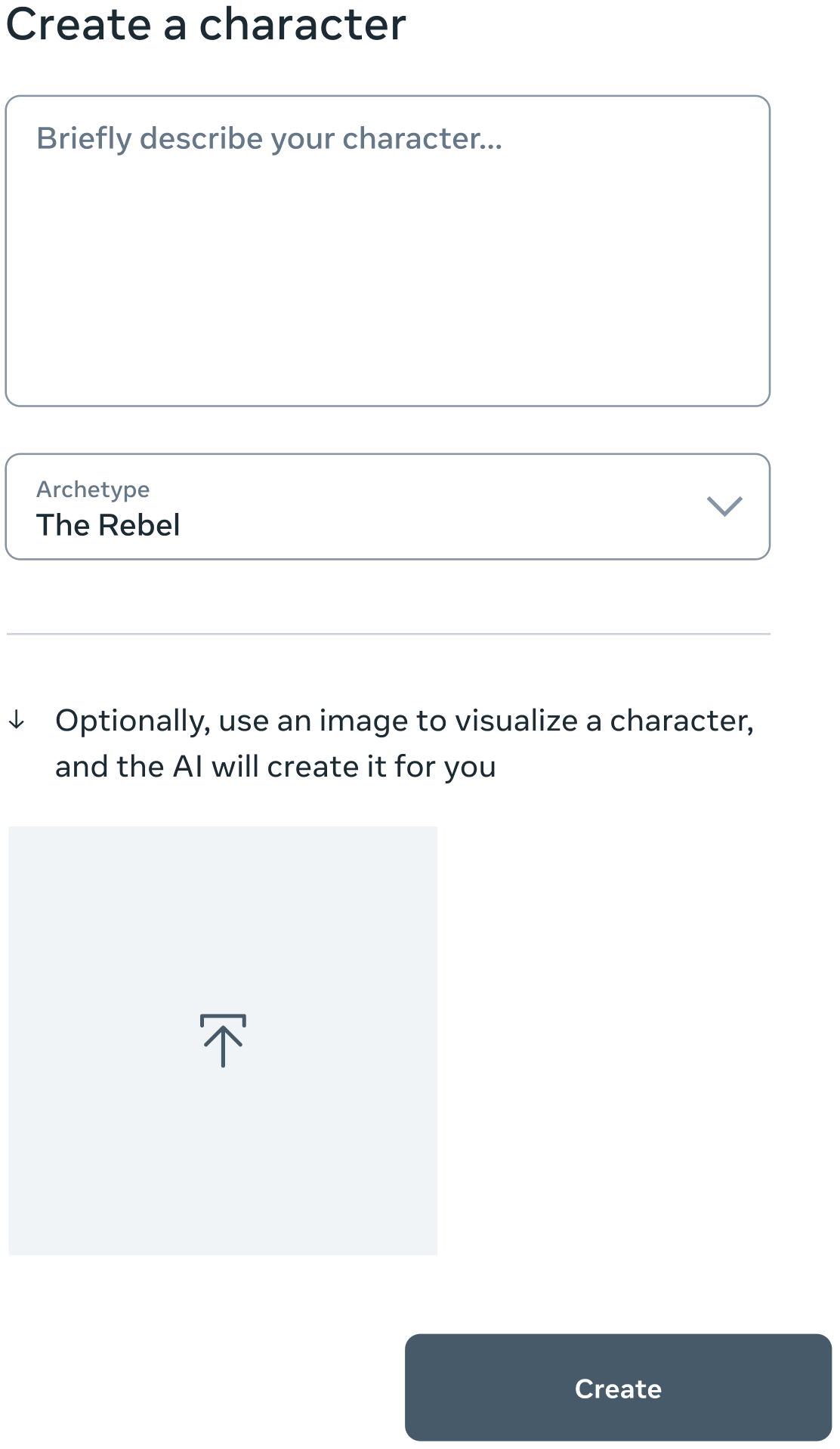}
         \caption{A suggested design idea to help the writer create a new character or refine an existing character idea.}
         \label{fig:1.2}
     \end{subfigure}

     \end{figure}
     \begin{figure}[!htb]
    \ContinuedFloat

     \begin{subfigure}{0.7\textwidth}
         \centering
         \includegraphics[width=\textwidth]{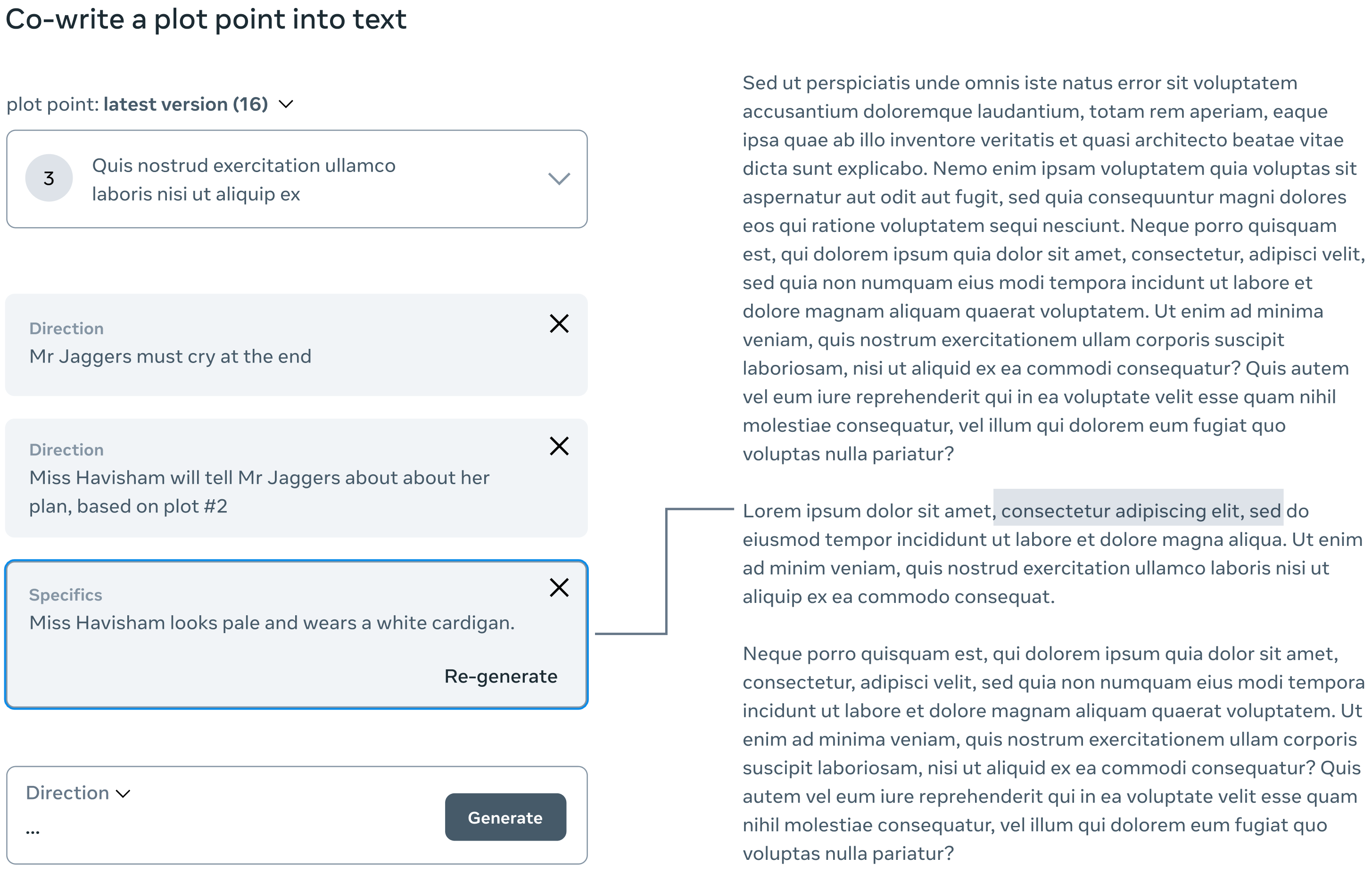}
         \caption{A design illustration of an interface helping the user receive help from AI on writing parts of the plot or text, each with functions or room for user direction.}
         \label{fig:1.3}
     \end{subfigure}
     
     \vspace{4em}
     
     \begin{subfigure}{0.4\textwidth}
         \centering
         \includegraphics[width=\textwidth]{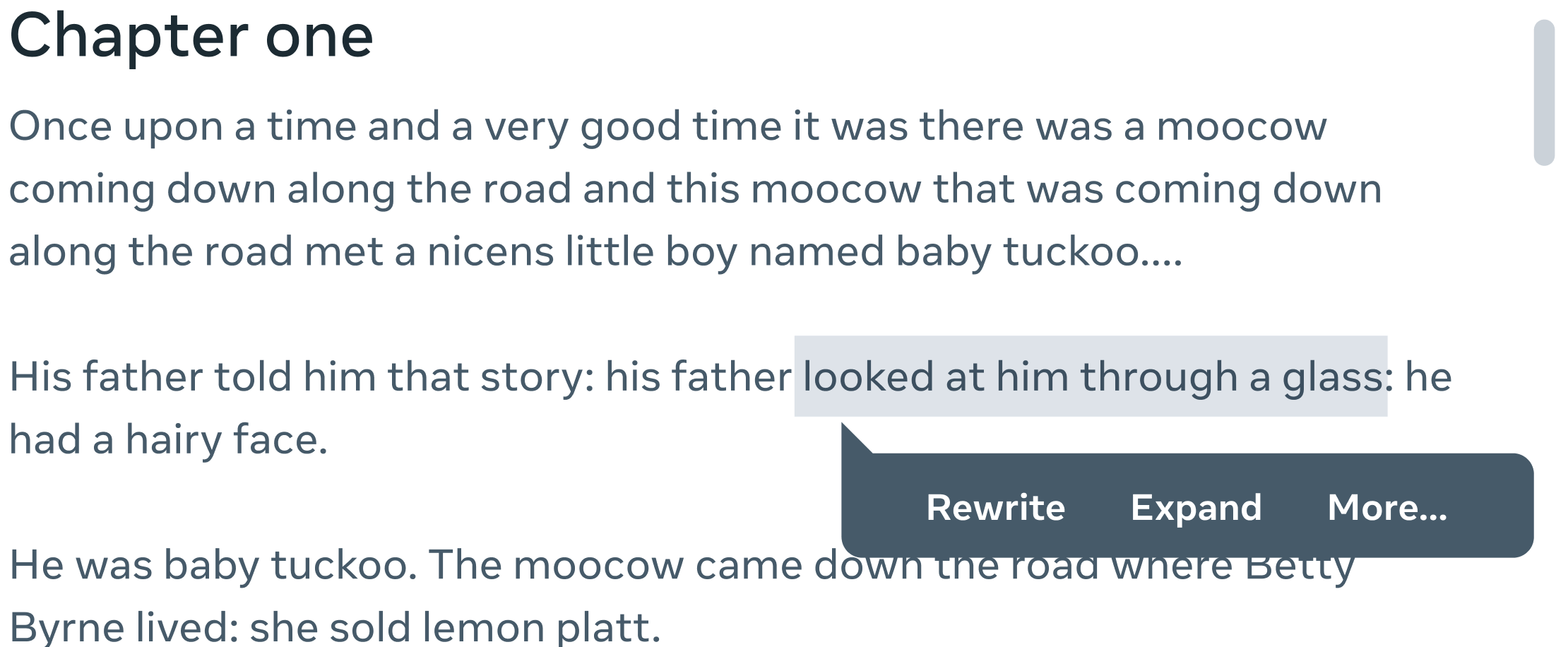}
         \caption{A simple illustration pointing out that AI can do more than write or rewrite, and specifically be given direction about what to do.}
         \label{fig:1.4}
     \end{subfigure}
     \hspace{2em}%
    \begin{subfigure}{0.5\textwidth}
         \centering
         \includegraphics[width=\textwidth]{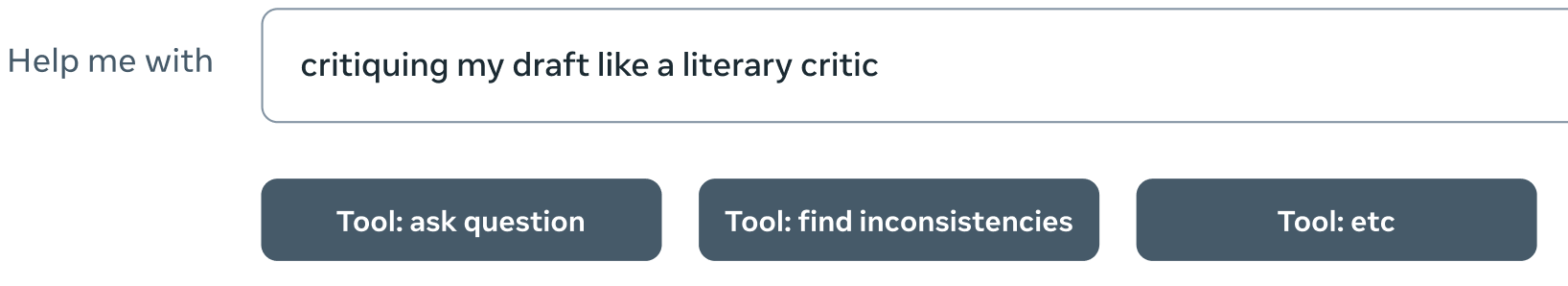}
         \caption{A design idea for an open-ended UI where the AI can suggest helpful actions too.}
         \label{fig:1.5}
     \end{subfigure}
        \caption{A selection of the various design illustrations used in our study, as stimuli, to help cause the formation of new ideas and provoking thoughts of writers. This set did not intend to lay out or imply a full interface, user journey, or exhaustive set of features.}
        \label{fig:mocks}
 \end{figure}

\end{document}